\documentclass[aps,amsmath,prl,reprint,showpacs,preprintnumbers,twocolumn]{revtex4}%,twocolumn]{revtex4}
\usepackage{amsfonts}
\usepackage{graphicx}
\usepackage{dcolumn}
 \usepackage{pdfsync}
\usepackage{multirow}
\usepackage{subfigure}
\usepackage{palatino}
\usepackage{graphicx}
\usepackage{epsfig}
\usepackage{ifpdf}
\ifpdf \DeclareGraphicsExtensions{.pdf,.png,.jpg,.mps}
 \else
\DeclareGraphicsExtensions{.eps} \fi

\begin{document}
\title{Efficiency and Its Bounds for Thermal Engines at Maximum Power using a Newton's Law of Cooling}
\author{H. Yan}
\affiliation{Department of Physics,
Indiana University/IUCF, 2401 Milo B. Sampson Lane, Bloomington, IN
47408, USA}
\email{haiyan@umail.iu.edu}
\author{Hao Guo}

\affiliation{Department of Physics, the University of Hong Kong}
\email{guohao.ph@gmail.com}

\begin{abstract}
We study a thermal engine model for which Newton's cooling law is obeyed during heat transfer processes. The thermal efficiency and its bounds at maximum output power are derived and discussed. This model, though quite simple, can be applied not only to Carnot engines but also to four other types of engines.
 For the long thermal contact time limit, new bounds, tighter than what were known before, are obtained. In this case,this model can simulate
 Otto, Joule Brayton, Diesel, and Atkinson engines. While in the
short contact time limit, which corresponds to the Carnot cycle, the same efficiency bounds as Esposito et al's  are derived.
In both cases, the thermal efficiency decreases as the ratio between the heat capacities of the working medium during heating and cooling stages increases.
 This might provide instructions for designing real engines.

\end{abstract}
\maketitle

\section{Introduction}
It is well known that real thermal engines can not achieve a perfect Carnot cycle. In a perfect Carnot cycle,the two reversible isothermal stages must be infinitely long and hence the Carnot engine has zero power output. Although the Carnot thermal machine is impractical, it gives an upper limit on the efficiency of all thermal engines. Real thermal engines work at finite cycle times and lose a finite amount of energy due to irreversible cycles and other mechanisms such as mechanical friction, heat leak and dissipative processes, etc. Searching for real thermal engines which operate with optimal cycles has caught a lot of attention. Here "optimal" refers to different optimizations of the heat engine, such as maximum efficiency, maximum power, maximum entropy production \cite{Salamon81} and maximum work\cite{Ondrechen83}, etc. Of all these optimizations, the efficiency of thermal engines at maximum output power is a very practical problem and has been extensively studied in the literature \cite{ESP10,CUR75,ESP09,BRO05}. The efficiency of a quantum thermal engine operating at maximum power has also recently been studied \cite{ABE11}.

One of the most important results addressing the efficiency of a thermal engine at maximum power was given by Curzon and Ahlborn in 1975\cite{CUR75}. Here we briefly review their result first. They made the assumption that during the time that the working medium is in contact with the hot(cold) reservoir, the amount of heat exchanged is proportional to the temperature difference between the working medium and the reservoirs, and also to the time duration of the processes. During the heating process, which lasts time $t_1$, the amount of heat $W_1$ absorbed by the system is
\begin{equation}\label{eq.w1}
W_{1}=k_1 t_{1}(T_{1}-T_{1w}),
\end{equation}
where $T_1$ is the temperature of the heat source, $T_{1w}$ the temperature of the working medium and $k_1$ the heat transfer coefficient of the heating process. Similarly, for the cooling process which lasts time $t_2$, the working medium releases heat $W_2$
\begin{equation}\label{eq.w2}
W_{2}=k_2 t_{2}(T_{2w}-T_{2}).
\end{equation}
Here $T_2$ is the temperature of the cold source, $T_{2w}$ the temperature of the working substance and $k_2$ the heat transfer coefficient of the cooling process. The reversibility of the adiabatic stages requires
\begin{equation}\label{eq.r1}
\frac{W_{1}}{T_{1w}}=\frac{W_{2}}{T_{2w}}.
\end{equation}
This leads to a relationship between $t_{1}$ and $t_{2}$. By maximizing the power output of the system, they derived the famous Curzon-Ahlborn (CA) formula for the efficiency of the thermal engines at maximum power as:
\begin{equation}
\eta_{CA}=1-\sqrt{\frac{T_{c}}{T_{h}}}.
\end{equation}
The CA formula describes the thermal engines of power plants very well \cite{ESP10,CUR75} and all the parameters here have clear physical meanings.
However, as pointed out by Ref. \cite{ESP10}, the CA formula is neither exact nor universal, and it gives neither an upper bound nor a lower bound.

In Ref. \cite{ESP10}, the authors considered a Carnot thermal engine performing finite-time cycles. They assume that the amount of heat absorbed by the system per cycle from the hot(cold) reservoir is given by
\begin{equation}\label{eq.1}
Q_{h}=T_{h}(\Delta S-\frac{\Sigma_{h}}{\tau_{h}}+...),
\end{equation}
and
\begin{equation}\label{eq.2}
Q_{c}=T_{c}(-\Delta S-\frac{\Sigma_{c}}{\tau_{c}}+...),
\end{equation}
where $T_{h,c}$ is the temperature of the hot(cold) reservoir and $\tau_{h,c}$ the time during which the thermal machine is in contact with the hot(cold) reservoir. The second terms of Eqs.(\ref{eq.1}) and (\ref{eq.2}) give the extra entropy production per cycle when the system deviates from the reversible regime. By maximizing the power, the efficiency of the engine can be derived. The upper and lower bounds of the thermal efficiency at maximum power are derived when the ratio $\Sigma_{h}/\Sigma_{c}$ approaches 0 and $\infty$ respectively. Esposito et al's result agrees well with the observed efficiencies of thermal plants\cite{ESP10,PHY}. However,  why the working
medium releases less heat for longer contact times with the cold reservoir as indicated by Eq.(\ref{eq.2}) was not explained. Also, in both Ref.\cite{ESP10} and \cite{CUR75}, only Carnot engines were studied.

In this paper,we study a more general and realistic thermal engine model and derive its efficiency bounds at maximum power. In both Ref.\cite{ESP10} and \cite{CUR75}, the temperature of the working medium does not change during heat transferring processes which is not true for either a realistic system, such as thermal plants, or for other heat engine models, such as the Otto, Joule-Brayton,Diesel, and Akinson engines\cite{Leff87}.
 Instead, we simply assume that heat transfer by a thermal engine is  described by Newton's law of cooling, thus it does not have to be isothermal anymore.
 Furthermore, we also take into account the fact that the thermal capacities of the working medium in
  realistic systems usually could be quite different at high and low temperatures\cite{KAV}. This is also be motivated by
  heat engines, such as the Diesel and Akinson type, for which the thermal capacities are different at the two different thermal stages\cite{Leff87}.

   With these two modifications, we argue that our model is not only more realistic, but also more general. Since the efficiency and its bounds
   are derived by considering heat exchange processes during which the temperature of the working medium could be close to or far away from
   isothermal, our model could simulate heat transferring not only in Carnot engines, as in Ref.\cite{ESP10} and \cite{CUR75}, but also
   some other engines such as Otto, Joule-Brayton, Diesel, and Akinson as described in Ref.\cite{Leff87}.

    The organization of the paper is as follows, we will first describe Newton's law of cooling and derive the corresponding
    entropy and heat formulas, and then study the thermal efficiency at maximum power for two limiting cases.

\section{Heat transfer and entropy production based on Newton's law of cooling}
We assume heat transferred by thermal engines in contact with a heat source is described by Newton's law of cooling:
\begin{equation}\label{eq.3}
\frac{dQ}{dt}=cm\frac{dT}{dt}=hA(T_{s}-T),
\end{equation}
where $c$ is the heat capacity, $m$ medium mass, $T$ medium temperature, $T_{s}$ heat source temperature, $h$ heat transfer coefficient,
and $A$ contact area. For convenience, we denote $hA$ by $k$. Though Newton's law of cooling is quite simple, many other heat transfer laws can be simplified to
it if the temperatures of the objects are high
 while the temperature difference between them is small. Based on this assumption, we consider a thermal engine working between hot and cold reservoirs at temperatures
  $T_{h}$ and $T_{c}$ respectively, and the initial temperature of the working medium is $T_{h0}$($T_{c0}$) at the beginning of the heating(cooling) stage.
   The solution to Eq.(\ref{eq.3}) gives the temperature of the working medium at time $t$:
\begin{equation}
T(t)=T_{h}+(T_{h0}-T_{h})e^{-\frac{k_{h}t}{c_{h}m}}=T_{h}+(T_{h0}-T_{h})e^{-\frac{t}{\Sigma_{h}}},
\end{equation}
where $\Sigma_{h}={c_{h}m}/{k_{h}}$.
Assuming that the time during which the working medium is in contact with the high temperature source is $\tau_{h}$, the entropy produced
 during the heating process can be evaluated straightforwardly as:
\begin{eqnarray}
\Delta S_{h}=-\int^{\tau_{h}}_{0}\frac{dQ}{T(t)}=c_{h}m\ln{\frac{T_{h}-xe^{-\frac{\tau_{h}}{\Sigma_{h}}}}{T_{h0}}},
\end{eqnarray}
where $x=T_{h}-T_{h0}$. The heat exchanged between the working medium and the high temperature source is given by
\begin{equation}\label{eq.10}
Q_{h}=\int_{0}^{\tau_{h}}k_{h}(T_{h}-T_{h0})e^{-\frac{t}{\Sigma_{h}}}dt=c_{h}mx(1-e^{-\frac{\tau_{h}}{\Sigma_{h}}}).
\end{equation}
Here and from now on we take
the convention that $Q>0$ means absorbing and $Q<0$ releasing heat. Similarly, the entropy production and heat exchange of the working medium during the cooling process are given by
\begin{eqnarray}
\Delta S_{c}&=&c_{c}m\ln{\frac{T_{c}+ye^{-\frac{\tau_{c}}{\Sigma_{c}}}}{T_{c0}}},\\
Q_{c}&=&-c_{c}my(1-e^{-\frac{\tau_{c}}{\Sigma_{c}}})\label{eq.12},
\end{eqnarray}
where $y=T_{c0}-T_{c}$ and $\Sigma_{c}={c_{c}m}/{k_{c}}$. After a thermodynamic cycle, the system returns to its initial state, and the total entropy change of the working medium should be zero $\Delta S_{h}+\Delta S_{c}=0$ \cite{CUR75}, which leads to
\begin{equation}\label{eq.4}
\ln{[(\frac{T_{c}+ye^{-\frac{\tau_{c}}{\Sigma_{c}}}}{T_{c0}})^{c_{c}}(\frac{T_{h}-xe^{-\frac{\tau_{h}}{\Sigma_{h}}}}{T_{h0}})^{c_{h}}]}=0.
\end{equation}
By noting that $T_{h0}=T_{h}-x$, $T_{c0}=T_{c}+y$ and by defining $\gamma\equiv c_{h}/c_{c}$, Eq.(\ref{eq.4}) is reduced to
\begin{equation}\label{eq.18}
(\frac{T_{c}+ye^{-\frac{\tau_{c}}{\Sigma_{c}}}}{T_{c}+y})(\frac{T_{h}-xe^{-\frac{\tau_{h}}{\Sigma_{h}}}}{T_{h}-x})^{\gamma}=1.
\end{equation}
The power output and the efficiency of the thermal engine are given by
\begin{eqnarray}\label{eq.19}
P&=&\frac{Q_{h}+Q_{c}}{\tau_{h}+\tau_{c}},\\
\eta&=&1+\frac{Q_c}{Q_h}.
\end{eqnarray}
 Generally, the efficiency $\eta_{m}$ at maximum power output can be derived using the constraint of Eq.(\ref{eq.18}). However, Eq.(\ref{eq.18}) is a transcendental equation which can not be solved analytically.
  In what follows we will focus our discussions on two special cases.

\section{Efficiency and its bounds in two special cases}
\subsection{Case I: Long contact time limit: $\tau/\Sigma\rightarrow \infty$}
\begin{figure}[htb]
\centering
  \includegraphics[clip,width=3.0in]{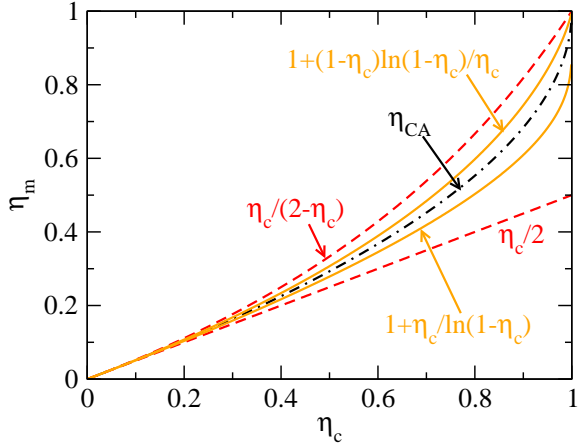}
  \caption{\small{(Color online) A comparison of upper and lower bounds for the long contact time limit between this work and \cite{ESP10}
  . The black dot-dashed line denotes the CA efficiency. The red dashed lines denote the upper and lower bounds of the thermal efficiency derived in Ref. \cite{ESP10}.
  The orange solid lines denote the bounds derived in this paper. }}
  \label{fig.scattering}
\end{figure}
In this case, the contact time is long enough that the working medium can exchange heat sufficiently with the reservoirs. Therefore the final temperature of the working medium
is close to the heat reservoir and quite different from its initial temperature.
 Numerically, when $\tau/\Sigma\sim5$,
we have $|T-T_{h0}|/|T_{h}-T_{h0}|\sim 0.007$ and $|T-T_{c0}|/|T_{c}-T_{c0}|\sim 0.007$. Thus, when $\tau/\Sigma$ is sufficiently large, which is supposed to be the case
studied in Ref.\cite{ESP10},
$\exp{(-\tau/\Sigma)}$ can be safely ignored, and Eq.(\ref{eq.18}) is reduced to
\begin{equation}\label{eq.17}
(\frac{T_{c}}{T_{c}+y})(\frac{T_{h}}{T_{h}-x})^{\gamma}=1.
\end{equation}
By plugging Eq.(\ref{eq.17}) into Eq.(\ref{eq.10}) and (\ref{eq.12}), using Eq.(\ref{eq.19})then the output power is given by
\begin{equation}
P=m\frac{c_{h}x-c_{c}T_{c}[(\frac{T_{h}}{T_{h}-x})^{\gamma}-1]}{\tau_{h}+\tau_{c}}.
\end{equation}
Let $\partial P/\partial x=0$, $P$ is maximized when
\begin{equation}
x=T_{h}[1-(\frac{T_{c}}{T_{h}})^{\frac{1}{1+\gamma}}].
\end{equation}
Therefore, the efficiency at maximum power is given by
\begin{equation}
\eta_{m}=1-\frac{1}{\gamma}[\frac{1-\frac{T_{c}}{T_{h}}}{1-(\frac{T_{c}}{T_{h}})^{\frac{1}{1+\gamma}}}-1]=1-\frac{1}{\gamma}[\frac{\eta_{c}}{1-(1-\eta_{c})^{\frac{1}{1+\gamma}}}-1].
\end{equation}
From the above expression we see that $\eta_{m}$ decreases as $\gamma$ increases. For the symmetric dissipation in which $\gamma=1$, $\eta_{m}$ becomes
\begin{equation}
\eta_{m}=1-\sqrt{\frac{T_{c}}{T_{h}}}.
\end{equation}
Interestingly, the CA efficiency is recovered though the situation is quite different. Expanding $\eta_{m}$ in series of $\eta_c$, we have
\begin{equation}
\eta_{m}=\frac{\eta_{c}}{2}+\frac{1}{12}(1+\frac{1}{1+\gamma})\eta_{c}^{2}+\mathcal {O}(\eta_{c}^{3}).
\end{equation}
The coefficient of the second order term lies between $1/12$ and $1/6$, while in Ref.\cite{ESP10} this term is between $0$ and $1/4$ which indicates a tighter bound here
. The lower and upper bounds of $\eta_{m}$ in this case are given by
\begin{equation}\label{eq.b1}
1+\frac{\eta_{c}}{\ln{(1-\eta_{c})}}\leq\eta_{m}\leq1+\frac{(1-\eta_{c})\ln{(1-\eta_{c})}}{\eta_{c}}.
\end{equation}
A comparison of the upper and lower bounds for the long contact time limit between this work and results derived in Ref.\cite{ESP10} is shown as FIG.\ref{fig.scattering}.
We see that the limits derived here give much tighter bounds than those derived in Ref.\cite{ESP10}.

We emphasize again that our model does not only apply to Carnot engines. Since the final
temperature of the working medium after heat exchange can be quite different from its initial temperature, it
is not necessarily an isothermal process and thus the engine does not need to be a Carnot type engine. It can also simulate the engines described in Ref.\cite{Leff87}, if we take $c_{h}=c_{c}=c_{v}$,
it is the Otto engine, $c_{h}=c_{c}=c_{p}$, the Joule-Brayton engine, $ c_{h}=c_{p},c_{c}=c_{v}$, the diesel engine and
$c_{h}=c_{v},c_{c}=c_{p}$, the Akinson engine. We can recover all the thermal efficiencies at maximum power derived
in Ref.\cite{Leff87}. Correspondingly, the bounds derived in this section should apply to those four types of engines mentioned above
in practical conditions.
\subsection{Case II: Short contact time limit: $\tau/\Sigma\rightarrow0$}
In this case, the heating and cooling processes are both short. Therefore the final temperature of the working medium after transferring heat  is very close to its initial temperature. This is approximately what was studied in Ref.\cite{CUR75} where temperature of the working medium does not change during heat transfers.
 Numerically, one can estimate that if $\tau/\Sigma\sim0.1$, then $|T-T_{h0}|/|T_{h}-T_{h0}|\sim 0.9$ and $|T-T_{c0}|/|T_{c}-T_{c0}|\sim 0.9$.
We solve Eq.(\ref{eq.18}) by expanding it as series of the infinitesimal variable ($\tau/\Sigma$) and matching both sides of the equation order by order (We always keep the same order of $\tau_{h}/\Sigma_{h}$ and $\tau_{c}/\Sigma_{c}$).
As $\tau/\Sigma\rightarrow0$, to the zeroth order of $\tau/\Sigma$, Eq.(\ref{eq.18}) simply gives $1=1$ which is trivial.
To the first order of $\tau/\Sigma$, Eq.(\ref{eq.18}) gives
\begin{equation}\label{eq.20}
\frac{\gamma x\frac{\tau_{h}}{\Sigma_{h}}}{T_{h}-x}=\frac{y\frac{\tau_{c}}{\Sigma_{c}}}{T_{c}+y}.
\end{equation}
Now the amounts of heat exchanged by the system during the heating and cooling processes are given by
\begin{eqnarray}
Q_{h}&=&c_{h}mx\frac{\tau_{h}}{\Sigma_{h}},\label{eq.Qh}\\
Q_{c}&=&-c_{c}my\frac{\tau_{c}}{\Sigma_{c}}.\label{eq.Qc}
\end{eqnarray}
The above equations agree with the fundamental equations listed at the beginning of Ref. \cite{CUR75}. Thus if we continue our straightforward calculation, we simply recover the same results in Ref. \cite{CUR75}, including the CA efficiency.
Now, we continue to expand Eq.(\ref{eq.18}) to the second order of $\tau/\Sigma$, we obtain another simple relation
\begin{equation}\label{eq.23}
\frac{y}{x}=\gamma\frac{T_{c}}{T_{h}}
\end{equation}
Combining this with Eq.(\ref{eq.20}), we get
\begin{equation}\label{eq.24}
\frac{\tau_{c}}{\Sigma_{c}}=\frac{T_{h}+\gamma x}{T_{h}-x}\frac{\tau_{h}}{\Sigma_{h}}
\end{equation}
To obatian the expression for the power output $P$, plug Eqs.(\ref{eq.23}) and (\ref{eq.24}) into Eq.(\ref{eq.19}), and expand the expressions of $Q_{h}$ and $Q_{c}$ to the first order of $\tau/\Sigma$ again, we have
\begin{equation}
P=\frac{c_{h}mx(1-\frac{T_{c}(T_{h}+\gamma x)}{T_{h}(T_{h}-x)})}{\Sigma_{h}+\frac{(T_{h}+\gamma x)\Sigma_{c}}{T_{h}-x}}.
\end{equation}
$P$ is maximized by letting $\partial P/\partial x=0$. Note $0<x<T_{h}$ and $x<\gamma T_{c}$, the unique allowed solution of $x$ is given by
\begin{equation}
x=\frac{[\sqrt{\frac{(T_{c}\Sigma_{h}+T_{h}\Sigma_{c})(1+\gamma)}{(T_{h}+\gamma T_{c})(\Sigma_{h}+\Sigma_{c})}}-1]T_{h}}{\frac{\gamma\Sigma_{c}-\Sigma_{h}}{\Sigma_{h}+\Sigma_{c}}}.
\end{equation}
Therefore the thermal efficiency at maximum power is given by
\begin{eqnarray}
\eta_{m}=1-\frac{T_{c}}{T_{h}}\frac{\frac{\gamma\Sigma_{c}-\Sigma_{h}}{\Sigma_{h}+\Sigma_{c}}-\gamma+\gamma\sqrt{\frac{(T_{c}\Sigma_{h}+T_{h}\Sigma_{c})(1+\gamma)}{(T_{h}+\gamma T_{c})(\Sigma_{h}+\Sigma_{c})}}}{\frac{\gamma\Sigma_{c}-\Sigma_{h}}{\Sigma_{h}+\Sigma_{c}}+1-\sqrt{\frac{(T_{c}\Sigma_{h}+T_{h}\Sigma_{c})(1+\gamma)}{(T_{h}+\gamma T_{c})(\Sigma_{h}+\Sigma_{c})}}}.
\end{eqnarray}
\begin{figure}[htb]
\centering
  \includegraphics[clip,width=3.0in]{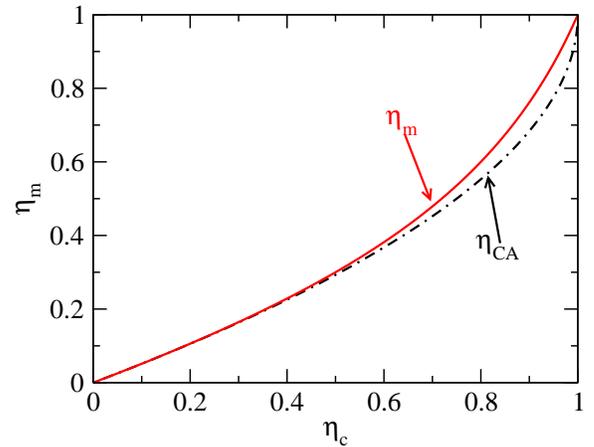}
  \caption{\small{(Color online) A comparison of Eq.(\ref{eq.cy}) (red solid line) with CA efficiency (black dot-dashed line). }}
  \label{fig.1}
\end{figure}
By defining $\beta=k_{h}/k_{c}$, and using the relation $\eta_{c}=1-{T_{c}}/T_{h}$,
$\eta_{m}$ can be expressed as
\begin{widetext}
\begin{equation}\label{eq.30}
\eta_{m}=1-\gamma(1-\eta_{c})\frac{\sqrt{[\beta+\gamma(1-\eta_{c})](\beta+\gamma)}-\sqrt{[1+\gamma(1-\eta_{c})](1+\gamma)}}{\beta\sqrt{[1+\gamma(1-\eta_{c})](1+\gamma)}-\sqrt{[\beta+\gamma(1-\eta_{c})](\beta+\gamma)}}.
\end{equation}
\end{widetext}
\begin{widetext}
\begin{equation}\label{E8}
\eta_m=1-\frac{\gamma^2_1}{\gamma^2_2}\frac{\gamma^2_2(1-\eta'_c)-1+\frac{1}{\gamma_1}\sqrt{\big(1+\gamma^2_1(1-\eta'_c)^2\big)(\gamma^2_1+\gamma^4_2)}}{\gamma^2_2+\gamma^2_1(1-\eta'_c)},
\end{equation}
\end{widetext}
Moreover, $\eta_{m}$ can be expanded in a series of $\eta_{c}$ as
\begin{equation}
\eta_{m}=\frac{1}{2}\eta_{c}+\frac{1}{8}(\frac{1}{1+{\gamma}/{\beta}}+\frac{1}{1+\gamma})\eta_{c}^{2}+\mathcal {O}(\eta_{c}^{3}).
\end{equation}
The coefficient of the first order term of $\eta_{m}$ is $1/2$, and the coefficient of the second order term lies in the range between $0$ and $1/4$. In the symmetric case where $\beta=1$ and $\gamma=1$, we have
\begin{equation}\label{eq.cy}
\eta_{m}=\frac{\eta_{c}(2-\eta_{c})}{4-3\eta_{c}}.
\end{equation}
When expanding as a series in $\eta_{c}$, the coefficient of the second order term is $1/8$. Those results agree with the expansion of CA efficiency \cite{ESP10}.
 To the third order of $\eta_c$,  the difference of $\eta_{m}$ from $\eta_{\textrm{CA}}$ is  $\eta_{c}^{3}/32+\mathcal{O}(\eta_{c}^{4})$. A comparison between $\eta_m$ and the
 CA efficiency and our result for the
 symmetric case is shown in FIG.\ref{fig.1}.

 Now, we estimate the bounds of $\eta_{m}$. In the limits $\gamma=0$ or $\gamma=\infty$ while $\beta$ is finite,
we recover the lower and upper limits of $\eta_{m}$ given in Ref. \cite{ESP10}.
\begin{equation}
\frac{\eta_{c}}{2}\leq\eta_{m}\leq\frac{\eta_{c}}{2-\eta_{c}}.
\end{equation}
Interestingly our bounds on $\eta_m$ are obtained in the short contact time limit ($\tau\rightarrow0$) while the same
results were obtained in the long contact time limit ($\tau\rightarrow\infty$) in Ref. \cite{ESP10}.
It is easy to verify that $\eta_{m}$ decreases as $\gamma$ increases, but increases as $\beta$ increases. This means the larger the ratio between heat capacities of the working medium at the hot and cold reservoirs, the lower the efficiency at maximum output power.

\section{Conclusions}
In summary, we presented an analysis of thermal efficiency and its bounds at maximum power for thermal engines for which the heat transferring processes
are described by Newton's law of cooling. In the long contact time limit, CA efficiency is recovered for symmetric thermal capacity and two tighter bounds
on the thermal efficiency are derived. The model can simulate Otto, Joule Brayton, Diesel and Atkinson engines in the long contact time limit.
In the short contact time limit, we recover the famous CA efficiency in the first order calculation. When we proceed to the second order calculation, we derived a different efficiency formula and recovered the efficiency bounds at maximum power given by Espositi, et al. In both limits, the thermal efficiency is found to decrease as $\gamma=c_{h}/c_{c}$ increases.This might be helpful for choosing a suitable working medium and working temperatures when designing a thermal engine whose heat transfer can be approximated by Newton's law of cooling. Other cases such as those associated with intermediate thermal contact time and different heat transfer laws are being investigated further.

This work was supported by U.S. Department of Energy, Office of Science under grant DE-FG02-03ER46093.
H.Yan thanks professor M.W. Snow for support. We thank Dr. Changbo Fu, E.Smith and Zhaowen Tang for stimulating discussions.
One of the referees for the previous version of this paper provided us very instructive suggestions and valuable
references, we acknowledge it and thank him or her.

\end{document}